\documentclass[12pt,aps,prb,preprint]{revtex4}   

\usepackage{graphicx} 
\usepackage{dcolumn}  
\usepackage{bm}       

\newcommand{\be}{\begin{equation}}
\newcommand{\ee}{\end{equation}}
\newcommand{\bea}{\begin{eqnarray}}
\newcommand{\eea}{\end{eqnarray}}

\newcommand{\al}{\mbox{$\alpha$}}
\newcommand{\te}{\mbox{$\theta$}}

\newcommand{\nn}{\mbox{$\nonumber$}}

\newcommand{\w}{\mbox{$\omega$}}

\begin{document}

\title{On the Flight of the American Football}

\author{C. Horn and H. Fearn}

\affiliation{Physics Department, California State University Fullerton, \\
Fullerton CA 92834 USA. \\
email: curtishorn@alum.rpi.edu,  hfearn@fullerton.edu}

\vspace{1.5in}
\date{\today}


\begin{abstract}
In this paper we examine the detailed theory of the American football in flight, with spin
and air resistance included. We find the theory has much in common with the theory of a gyroscope
and also rocket trajectory with a misaligned thruster. Unfortunately most of the air resistance data,
for rocketry and ballistics,
is for speeds of Mach 1 or higher, where the air resistance increases dramatically.
We shall approximate a realistic air resistance, at the slower speeds of football flight,
with a drag force proportional to cross sectional area and either $v$ or $v^2$, depending on speed,
where $v$ is velocity of the football.
We begin with a discussion of the motion, giving as much detail as possible without the use
of complex analytic calculations. We point out the previous errors made with moments of inertia and
make the necessary corrections for more accurate results.
\end{abstract}



\maketitle


\section{Introduction}

\noindent
It has come to our attention that there are still some unresolved problems
relating to football flight. Dr. Timothy Gay, author of a popular book on
Football Physics \cite{tim1} suggested a challenge at a recent meeting of the Division
of Atomic, Molecular and Optical Physics, in Tennessee, May 2006.
The challenge given at the conference was to explain why the long axis of the ball
appears to turn (pitch) and follow the parabolic trajectory of the flight path.
After an extensive literature search we have found that the theory of why a football
pitches in flight has been explained quite well by Brancazio \cite{bran1}.
The more general theory of a football in flight, giving elaborate mathematical details
has been given by both Brancazio and Rae \cite{bran2, bran3, rae1}.
We have found that the literature (especially comments found online) do tend to use
improper moments of inertia for the football and incorrectly apply the Magnus force
(known as spin drift in ballistics) to
explain yaw of the football in flight. Clearly, since a challenge was posed by an expert
in football physics, we feel the need to summarize the literature in this area
and update it where necessary.\\

We intend to give the reader all the mathematical details needed for an accurate
description of the moments of inertia for the football. The needed theory, on torque free precessional
motion (for more than one spin axis) and gyroscopic motion when torque is applied,
is already available in standard mechanics texts \cite{goldstein, marion}.
With the use of these texts and papers like that
of Brancazio \cite{bran1, bran3} we take the theory to be well enough expounded. We will merely
quote the needed equations and cite the references.\\

The second author would also
like to take this opportunity to write a paper on ``something typically American" to
celebrate becoming an American citizen in June 2006.\\

Several scientists at SUNY; State University of New York at Buffalo, have an e-print online
with flight data taken with an onboard data recorder for an American football.
They used a ``Nerf" ball and cut out the foam to incorporate circuit boards and four
accelerometers inside the ball \cite{nowak}. They confirm that a typical spin rate is
10 rev/sec or 600 RPM (revolutions per minute) from their data. They also give a precession frequency of
340 RPM which gives a 1.8 value for the
spin to precession rate ratio (spin to wobble ratio).
We will discuss this further later on.

The usual physics description of football flight, of the kind you can find online,
is at the level of basic center of mass
trajectory, nothing more complex than that \cite{hsw}. It is our
intention in this article to go into somewhat greater detail in order
to discuss air resistance, spin, pitch and yaw. We find that the literature uses, for the most part,
a bad description of the moments of inertia of the football. We intend to remedy that.
Our review will encompass all possible flight trajectories,
including kicking the ball and floating the ball.
 We will briefly review the equations needed for all possible scenarios.\\

\noindent
In order to conduct this analysis it will be necessary to make some basic assumptions. We assume
that solid footballs may be approximated by an ellipsoid, we ignore any effect of the laces. We note that
a Nerf ball (a foam ball with no laces) has a very similar flight pattern to a regular inflated football,
so the laces can be thought to have little or no effect, they just help in putting the spin on the ball.
More about this later. For game footballs we will consider a prolate spheroid shell and
a parabola of revolution for the shapes of the pigskin.
We calculate the moments of inertia (general details in Appendix A and B) and take the long axis
of the ball to be between 11--11.25 inches and the width of the ball to be between 6.25--6.8 inches.
These measurements correspond to the official size as outlined by the
Wilson football manufacturer \cite{wilson}. The variation must be due to how well the ball is inflated,
and it should be noted the balls are hand made so there is likely to be small variation in size
from one ball to the next.
(The game footballs are inflated to over 80 pounds pressure to ensure uniformity of shape,
by smoothing out the seams.)
In the following we give the description of the football in terms of
ellipsoidal axes, whose origin is the center of mass (COM) of the ball.

\noindent
\begin{figure}
  \includegraphics[width=3.0in]{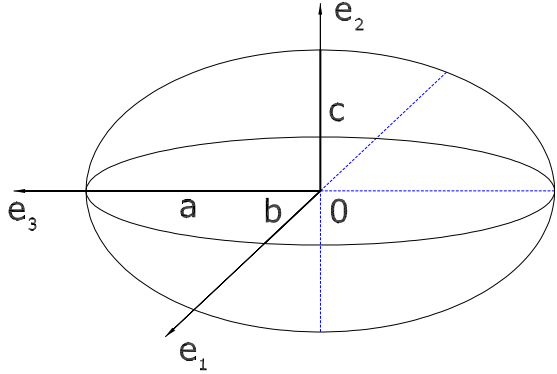}\\
  \caption{Basic ellipsoidal shape of football showing axes.}\label{Fig. 1.}
\end{figure}

\noindent
For explanation of the terms in an ellipse see references \cite{goldstein, marion}.
We take the semi--major axis to be $a$. The semi--minor axes are $b$ and $c$.
For the football, we will assume both the semi--minor axes are the same
size $b=c$ and they are in the transverse, $e_1$ and the $e_2$, unit vector directions.
The semi--major axis is in direction $e_3$. These are the principal axes of the football.\\

\noindent
We will take the smallest length of an official size football,
$11$ inches and largest width of $6.8$ inches, we get $a = 5.5$ inches (or $14.1 cm $) and
$b=3.4$ inches (or $8.6 cm$) which gives a ratio $a/b = 1.618$ which agrees
with the average values given by Brancazio \cite{bran3}.
Using the solid ellipsoid model, this gives us the principal moments of inertia
$I_1 = I_2 = I = \frac{1}{5} m b^2 \left( 1  + a^2/b^2 \right) $
and $I_3 = 2 m b^2/5 $.\\

\noindent
The torque free spin the wobble ratio (or its inverse) can be found in most advanced text books on
gyroscopic motion or rigid body motion, \cite{marion}. We can give the formula here for the torque free
ratio of spin to wobble (or precession);
\be
\frac{ \mbox{ spin}}{\mbox{wobble}} = \frac{\w_3}{\dot{\phi}}= \frac{I \cos \theta }{I_3} =
\frac{1}{2} \left( 1 + \frac{a^2}{b^2} \right) \cos \theta
\ee

\noindent
Clearly, for a horizontal ball, the Euler angle $\theta =0$,
if we use the ratio of semi major to semi minor axis $ a/b = 1.618$ we get
spin/wobble $ = 1.603$, which is less than the experimentally observed value,
of $1.8$ \cite{nowak,tim1}. This corresponds to a vacuum
spin to wobble ratio, no air resistance has been taken into account. It appears that the precession
is somewhat effected by air drag. Also the shape may not be that accurate,
since an ellipsoid has fairly rounded ends and a football is more pointed.

\noindent
There are several available expressions for air resistance. In ballistic theory there is
the Prandtl expression for force of air resistance (or drag) \cite{marion} which goes
$f_d = 0.5 c_w \rho A v^2$, where
$c_w$ is some dimensionless drag coefficient (tabulated), $\rho$ is the density of air,
$A$ is the cross--sectional area and $v$ is the velocity of the projectile. This formula
is only valid for high velocities, of $24 $ m/s or above, up to Mach--1 where the air resistance
increases rapidly. (Mach--1 is the speed of sound, which is $330$ m/s or about $738$ mph. )
It turns out for lower speeds (less than $24 $ m/s or equivalently $54$ mph)
the Stokes' Law formula is more appropriate which has an air resistance proportional
to velocity not the square of the velocity. Generally, air resistance formulae that go with the square
of the velocity are called Newton's laws of resistance and those that go as the velocity
are called Stokes' Laws \cite{marion}.\\

\noindent
From the information online and in papers and books, we have discovered that the average
speed an NFL player can throw the ball is about $20$ m/s or $44.74$ mph \cite{watts}. The top speed
quoted by a football scout from a $ 260 \; \ell$ b quarterback was clocked at $56$ mph or $25$ m/s.
It is doubtful that any quarterback can throw faster than $60$ mph or $26.8$ m/s. It appears that these
velocities are right on the boarder line between using one form of air drag force and another.
If we assume that most quarterbacks will throw under less than ideal conditions, under stress, or
under the threat of imminent sacking, they will most likely throw at slightly lower speeds, especially
as the game continues and they get tired. We further suggest that it is easier to
throw the ball faster when you throw it in a straight line horizontally,
as opposed to an angle of 45 degrees above the horizontal for
maximum range. Therefore, we suggest that the air drag force law
should be of the Stokes variety, which is proportion to velocity and cross
sectional area of the ball and air density. We shall use an air resistance of
$f_d = \gamma \rho A v $ where $\rho $ is the density of air,
$A$ is the cross--sectional area of the football and $v$ is the velocity of the ball.
The $\gamma$ factor takes into account all other dimensions and can be altered to get
accurate results. We shall assume that an average throw reaches speeds of up to $20 $m/s or
$44.74$ mph. (The conversion factor is $0.44704 $ m/s $ = 1 $ mph, see \cite{iop} ).\\

\noindent
In rocket theory, and generally the theory of flight, there is a parameter known as the
center of pressure (CP) \cite{space}. The center of pressure of a rocket is the point
where the pressure forces act through. For example, the drag force due to air resistance would
be opposite in direction to the velocity of the ball and it would act through the CP point.
It is generally not coincident with the center of mass
due to fins or rocket boosters, wings, booms any number of ``attachments" on aeroplanes and rockets.
It is clear that if the CP is slightly offset (by distance $\ell$ say) from the COM of
the football this would lead to a
torque about the COM since torque is equal to the cross product of displacement and force,
$ \vec{\tau}=\vec{\ell} \times \vec{f_d }$.

\noindent

\begin{figure}
  \includegraphics[width=6.5in]{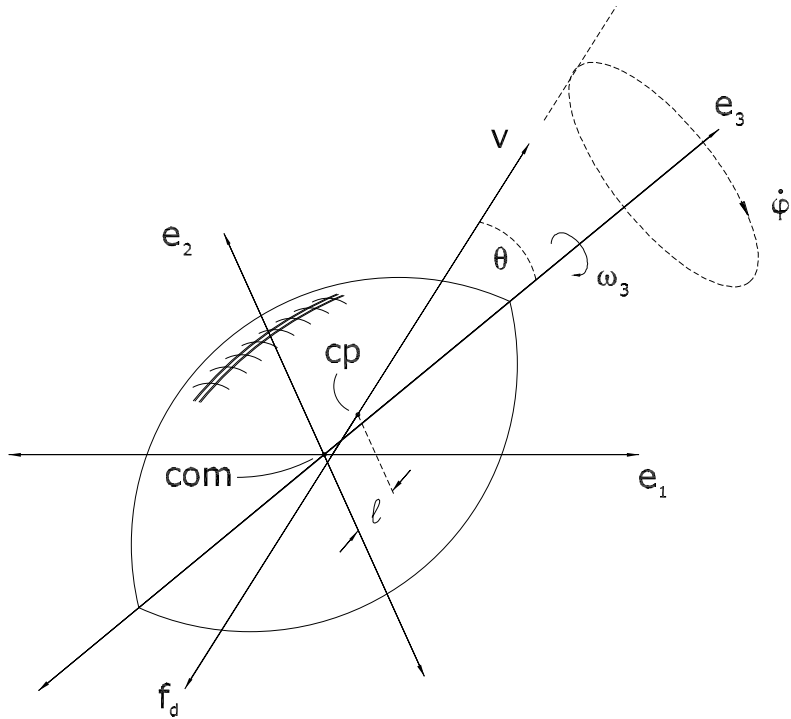}\\
  \caption{Center of mass, center of pressure (CP), velocity and drag force on the ball. The drag force
  $f_d$ acts along the velocity direction and through the CP. We take $\theta$ as the angle
  between the force and the $e_3$ axis. The aerodynamic torque is then $\tau = f_d \ell \sin \theta $.}
  \label{Fig. 2.}
\end{figure}

\noindent
The center of pressure was mentioned also by Brancazio \cite{bran1, bran2, bran3}
as an important factor in the flight of the football. In fact
the laces will tend to make the CP slightly offset from the COM which will add a small torque, but
this is negligible in comparison to other effects which we describe in the next section.
The offset of the CP from the COM is caused by aerodynamic drag on the leading
edge of the football in flight. This would offset the CP slightly forward of the COM.
 The CP offset results in gyroscopic precession
which in turn is due to torque since the drag forces act through the CP not the COM. A second
form of precession, called torque free precession, comes from the way the football is thrown.
If there is more than one spin axis the ball will precess in flight. \\

\noindent
With these definitions made, we are now ready to discuss the flight of the football.

\noindent
\section{Discussion of minimization of energy in flight}

\noindent
We wish to explain the pitching and possible yaw of the football during its flight with energy
minimization principals which have not been discussed previously in the literature.
The center of mass motion of the ball follows a perfect parabolic path, as explained in the
basic treatments already mentioned. We may treat the rotational motion separately, and consider the
body frame of the football, centered on the COM of the ball, as it moves.\\

We will assume that the ball is thrown, by a professional footballer, so that it has a ``perfect"
spin along the $e_3$ axis, see Fig. 1.
However, no matter how well the ball is thrown, there will always be
a slight torque imparted to the ball because of the downward pull of the fingers needed to
produce the spin. The effect of this ``finger action" is to tilt the top of the ball upward away
from the velocity or thrust direction, $\vec{v}$. Thus, the velocity vector $\vec{v}$ is not
exactly in the same direction
as the spin, $e_3$ axis. This misalignment of the initial throw (initial thrust vector not in
the $e_3$ direction) results in a torque about the $e_1$ axis. This produces a pitch of the ball
during its flight. This pitching is furthermore advantageous, since it tends to reduce the air
resistance of the football. We stated in the introduction that the drag was proportional to the
cross sectional area and the velocity of the football. If the ball is thrown at speed $v_0$ at an
angle to the horizontal of $\pi/4$ to maximize range, then initially the horizontal and vertical
velocities are the same. Both directions suffer air resistance but the vertical direction also
works against gravity so the vertical velocity will decrease faster than the horizontal.
The air drag is proportional to cross sectional area, so it is energetically favorable for the
football to pitch in flight and present a larger cross section vertically as its upward speed slows down.
When the ball is at maximum height, there is zero vertical velocity. The football can present
its maximum cross section vertically since this will not effect the air resistance vertically, which
depends on velocity. At this maximum height, the horizontal cross section is also a minimum.
This reduces the horizontal drag to a minimum which is again energetically favorable. One should note that
objects in nature tend to move along paths of least resistance and along paths which minimize energy.
This is exactly what the football will do. One should further note that the moments of inertia
about the $e_1$ and $e_2$ axes are the same and $1.655$ times larger (using experimental values) than the
moment of inertia about the spin axis $e_3$, for $a/b = 1.618$.
It is well known in space physics when a rotating body (a satellite in space
for example) has $2$ or $3$ different moments of inertia along different axes, and there is some
dissipation of energy, no matter how small, there is a natural tendency for kinetic energy to be
minimized, which results in a change of spin axis. Let us elaborate.
If there is no torque (gravity cannot exert a torque) then any initial angular
momentum is a conserved quantity. If there is some kind of energy dissipation, (in the case of the
football air resistance) then the kinetic energy will be minimized
by rotating the spin axis to the axis of largest moment of inertia. The kinetic energy is
$L^2/( 2I)$ where $L$ is angular momentum  and $I$ is the moment of inertia. Clearly the largest
moment of inertia minimizes energy. Since the $I$ in the $e_1$ and $e_2$ directions is the same,
why does the football not ``yaw", rotate about the $e_2$ axis? Well it turns out that the initial
throw does not initiate this torque (not if the ball is thrown perfectly anyway!) and this rotation
would act to increase the air drag in both the horizontal and vertical directions, so it is not
energetically favorable. \\

\noindent
For a non--perfect throw, which most throws are, there is a slight initial torque in
the $e_2$ direction also caused by the handedness of the player. A right handed player will
pull with the fingers on the right of the ball, a left handed player pulls to the left,
this will result in a yaw (lateral movement). This yaw will result in the football moving slightly
left upon release and then right for a righthanded player and slightly right upon release and then
left for a left handed player. This strange motion has been observed by Rae \cite{rae1}.
The lateral motion is a small effect
which would only be noticeable for a long (touchdown pass of approx 50 m) parabolic throw.
The effect is sometimes
attributed to the Magnus (or Robin's) force, which we believe is an error. The angular velocity of
the spin is not sufficiently high for the Magnus force to have any noticeable effect. We suggest
that the slight lateral motion is just a small torque imparted by the initial throw, due to the fingers
being placed on one side of the ball and slightly to the rear of the ball. One can imagine the initial
``thrust" CP being slightly off center (towards the fingers) and towards the rear of the ball. Hence the
resemblance to a faulty rocket thruster!
The force of the throw greatly exceeds the drag force hence the CP is momentarily closer
to the rear of the ball and slightly toward the fingers or laces. As soon as the football
is released the aerodynamic drag will kick in and the CP shifts forward of the COM.
During flight through the air the CP of the football moves forward because
now the aerodynamic drag is the only force acting through the CP which can produce a torque, and
the drag force is directed towards the leading edge of the football not the rear. For a right handed player,
the initial torque from the throw will send the football slightly to the left. This will switch
almost immediately upon release as the CP shifts forward of the COM and then the ball will move towards
the right of the player. The overall lateral motion will be to the right for a right handed player. This
fairly complex motion has been observed in practice by Rae \cite{rae1}.\\

\begin{figure}
  \includegraphics[width=6.0in]{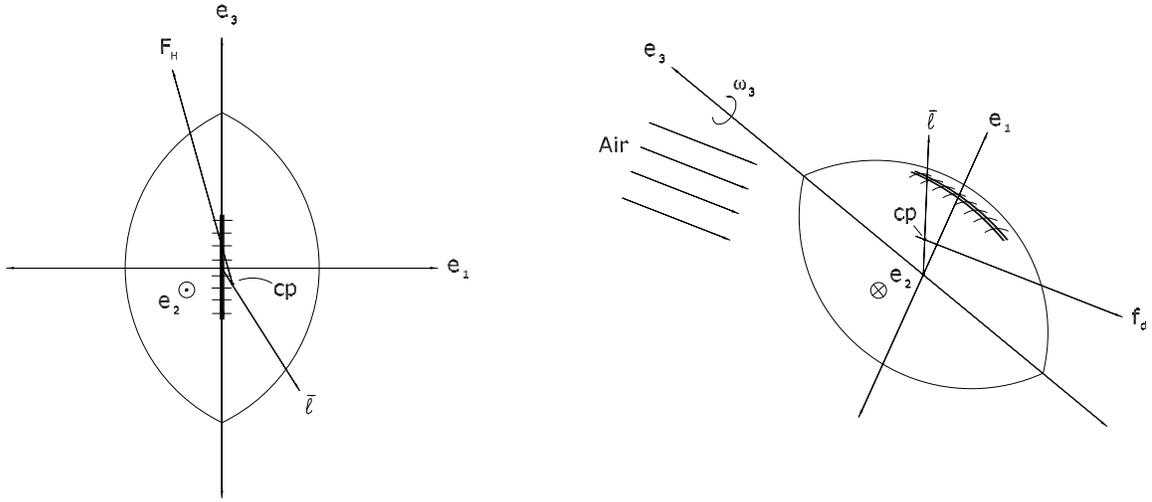}\\
  \caption{This figure shows the initial throw and the position of the CP towards the hand, and the CP in flight when it shifts forward.
  The torque directions are also shown as $e_2$. We assume the football is thrown by a right handed player. First it yaws left and then
  right. The overall motion is slightly to the right for a right handed player.}\label{Figure 3.}
\end{figure}

The Magnus force will be important for kicking the football, where the velocity of the ball and
the imparted spin is much greater. To give an example when the Magnus force is great, consider
a curve-ball in baseball. The typical velocity of the ball is $v= 80$ mph (35.76 m/s) the
mass $m= 0.145$ kg, the distance travelled is approximately $L=18$ m and the spin can be as high as
$\omega = 2000$ rpm (or 33.33 rev/sec ). The mass parameter $K_2 = 5.5 \times 10^{-4} $ kg. This
gives a Magnus deflection $d$ of \cite{ferrer}

\be
d = \frac{K_2 L^2 \omega}{2 m v} \;\;\; .
\ee

\noindent
For the curve-ball, $d= 0.57$ m. Now let us consider typical values for a football. Consider a pass
of length $L=50$ m. The football mass is $m=0.411$ kg, a throwing velocity of $v= 20$ m/s and a
spin of $\omega= 600 $ rpm ( equivalent to 10 rev/sec). These numbers give a Magnus deflection
of $d=0.8$ m for a 50 m pass. This would be hardly noticeable. A strong gust of wind is more likely to
blow the football off course than it is to cause a Magnus force on it. The only way to really account
for a right handed player throwing slightly to the right and a left handed player throwing slightly
to the left would be in the initial throw. This must have to do with finger positioning, and laces,
on one side of the ball or the other.

\section{The theory of flight}

\noindent
After an extensive search of the literature, we have discovered a detailed analysis
on the rigid dynamics of a football by Brancazio \cite{bran1,bran2, bran3}.
Clear observations of lateral motion, of the type discussed above,
have been made by Rae \cite{Rae1}.
The moments of inertia used are that for an ellipsoidal shell which have been calculated
by subtracting a larger solid ellipsoid  of semi--major and minor axes $a+t$ and $b+t$ from
a smaller one with axes $a$ and $b$. The results are kept in the first order of the
thickness of the shell, $t$. We have calculated the exact prolate spheroid shell moments (in Appendix A)
and we have also used a parabola of revolution, (calculation in Appendix B)
which we believe more closely fits the shape of a football.
Our results are seen to complement those of Brancazio \cite{bran3}.\\

\noindent
The angular momentum of a football can only change if there is an applied torque.
Aerodynamic drag on a football can produce the torque required to pitch
the ball to align the $e_3$ axis along the parabolic trajectory \cite{bran1}. The air
pressure over the leading surface of the football results in a drag force which acts through the
center of pressure (CP). The CP depends on the speed of the football and the inclination of
the $e_3$ axis to the horizontal. For inclinations of zero and 90 degrees there is no torque since the CP
is coincident with the COM, ignoring any effect of the laces. During flight when there is an acute angle
of inclination of the $e_3$ axis to the horizontal, the CP should be slightly forward of the COM, since
the air hits the leading edge of the ball. The gyroscopic precession of the ball is caused by this
aerodynamic torque. The resulting motion of the football is very similar to
a gyroscope \cite{bran1, goldstein, marion} but has extra complexity due to the drag forces
changing with pitch angle of the football. For stability we require that \cite{bran1},
\be
\omega_3 = \frac{ \sqrt{ 4 \tau I_1 }}{I_3} \;\;\; ,
\ee

\noindent
where $\tau = f_d \ell \sin \theta$ is aerodynamic torque, $\theta$ is the angle between
the aerodynamic drag force and the $e_3$ axis,
$I_1$and $I_3$ are the moments of inertia in the transverse and
long $e_3$ axis directions and $\omega_3$ is the angular velocity of the football about the $e_3$ axis.
The gyroscopic precession rate, defined as the angular rotation rate of the velocity axis
about the $e_3$ axis, $\dot{\phi}$ is given by \cite{bran3}
\be
\dot{\phi} = \frac{ f_d \ell}{I_3 \omega_3 }
\ee
where $f_d$ is the aerodynamic drag (which is the cause of torque $\tau$ above) and $\ell$ is
the distance from the CP to the COM. Both of these values will change with the pitch of the ball,
hence the football dynamics is rather more complex than that of a simple top.

\noindent
For low speeds of the ball, when the aerodynamic drag is small,
there can still be a precession of the football due to an imperfect throw. That is,
if there is more than one spin axis of the football it will precess in flight with no torque.
The ball gets most of its spin along the long $e_3$ axis. However, because the ball is held at one end,
a small spin is imparted about the transverse axis. A slight upward tilt of the ball on release
is almost unavoidable
because the fingers are pulling down on rear of the ball to produce the spin. Thus, there is an
initial moment about the $e_1$ axis which will tend to pitch the football. This non--zero spin
will result in torque--free precession or wobble.

\begin{figure}
  \includegraphics[width=6.0 in]{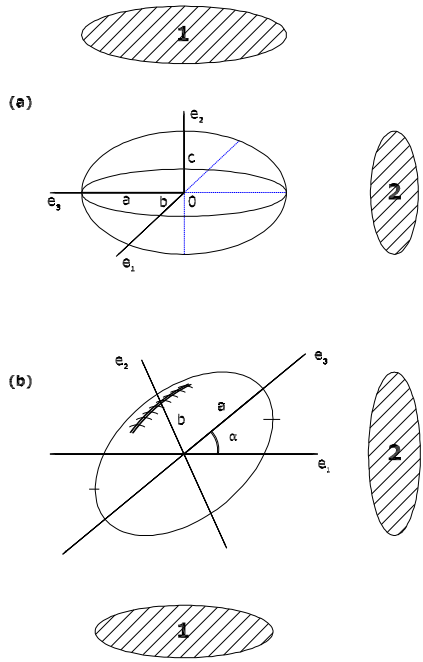}\\
  \caption{The surface area the football presents,
  at different inclination angles in flight. Fig 4(a) show maximum surface area $ \pi ab$ vertically
  and minimum surface area $ \pi b^2 $ horizontally. Fig. 4(b) has an inclination of $\alpha$ and the
  surface area this football presents to the vertical and horizontal has changed.}\label{Fig. 4.}
\end{figure}

\noindent The aerodynamic drag forces are linearly dependent on
the surface area $A$ of the football. The surface area $A$ would
be in the direction of motion. Figure 4(a) shows a football which
is perfectly horizontal, or with zero inclination angle. The
vertical surface area 1, it presents has a maximum area $\pi a b $
and the horizontal surface area 2 it presents has a minimum area $
\pi b^2$. Figure 4(b) shows a football at angle of inclination
$\alpha$. The surface area has now changed. The vertical surface
area 1 has become $ \pi b ( a \cos \alpha + b \sin \alpha )$ and
the horizontal surface has an area $\pi b ( a \sin \alpha + b \cos
\alpha)$. The velocity of the football can easily be transformed
into the vertical and horizontal components and thus the
aerodynamic drag $f_d$ can also be written in terms of vertical
and horizontal components for each angle $\alpha$. The equations
of motion are tedious to write out but a computer code can easily
be constructed to plot the football position and orientation in
flight. It is recommended that the parabola of rotation moments of
inertia be used as the most accurately fitting the football.

\section{Conclusions}

\noindent
It appears that footballs, have something in common with natural phenomenon, in that
they tend to follow the path of least resistance (air resistance) and the motion tends to
minimize energy. Also, when in doubt about projectile motion, ask a rocket scientist!

\noindent
The experimental values of the moments for the football, as determined
by Brody \cite{brody} using a torsion pendulum and measuring periods of oscillation are;
$I_1 = 0.00321 $kg $m^2$ and $I_3 = 0.00194 $kg $m^2$ and the ratio $I_3 / I_1 = 0.604 $.
Drag forces on a football have been measured in a wind tunnel by Watts and Moore \cite{watts}
and independently by Rae and Streit \cite{rae2}.

\noindent
For the prolate spheroid shell football we obtained the following moments of inertia,
(these results were checked numerically on Mathematica 5.2),
using $a=0.141 $ m (or $5.5$ in), $b=0.086$ m ( or $3.4$ in) and $M =0.411$ kg
were $I_1 = 0.003428 $ kg$m^2 $ and $I_3 = 0.002138$ kg$m^2$.
When we use the same parameters in our exact formulae (see Appendix A)
we find exactly the same numbers, so we are
confident that the above results are correct. \\

\noindent
For the parabola of revolution, we get $I_1 = 0.002829 $ kg$m^2$ and $I_3 = 0.001982$ kg$m^2$
(see Appendix B for details).
We suggest that the moment of inertia $I_1$ is slightly lower than the experimental value
because of extra leather at the ends due to the seams which we have not taken into account.
This is caused by the four leather panels being sewn together and there being a little
excess material at the ends making the football slightly heavier at the ends than we have accounted for.
If we add a small mass to either end of the football this would account for the very small
increase in the experimentally found value. The increase in moment of inertia required
(experimental value - our value)is
$\Delta I_1 = 0.000381$ kg$m^2$ which could be accounted for by a small mass of leather $m_0/2$ at
either end of the ball, where $\Delta I = m_0 a^2$ and $a$ is the semi-major axes $0.141 $ m. Hence,
$m_0 = 19.164 $ g (grams) which implies $m_0/2 = 9.582$ grams excess leather at each end of the ball.
This is a very small amount of leather! We believe this is a more accurate description of the
football than the prolate spheroid shell or the solid ellipsoid.\\

\noindent
Furthermore, the solid ellipsoid gives quite different moments.
For the solid, $I_1 = (1/5)m(a^2 + b^2) = 0.002242 $ kg$ m^2 $ , for the same $a,b$ as above
and $ I_3 = (2/5) m b^2 = 0.001216 $ kg$ m^2 $.\\

\section{Acknowledgments}

\noindent
We would like to thank Dr. M. Khakoo of CSU Fullerton for telling
us about the ``football" challenge set by Dr. Timothy Gay, at the
37th meeting of the Division of Atomic, Molecular and Optical Physics (DAMOP)
meeting in Knoxville Tennessee, May 16--20th 2006. Mr. Horn completed this project as part
of the requirements of a Masters degree in Physics at CSU Fullerton under the
supervision of Prof. H. Fearn.

\section{Appendix A}

Derivation of the principal moments of inertia of a prolate spheroidal shell (hollow structure).
The football is roughly the shape of a prolate spheroid, which is an ellipsoid with two semi
major axes the same length. The equation for the prolate spheroid is;
\be
\frac{x^2}{b^2} +\frac{y^2}{b^2} +\frac{z^2}{a^2}=1 \label{elli}
\ee
where $a$ is the semi major axis aligned along the length of the football. This will be the spin axis.
$b$ is the semi minor axis in both the $x$ and $y$ directions. We assume $a > b$. In fact for
an official size football we will take $a=5.5$ inches and $b= 3.4$ inches, this will be useful for
later numerical examples.
It is appropriate to introduce prolate spheroidal coordinates \cite{schaum},
to calculate the moments of inertia.
\bea
x &=& \al \sinh \varepsilon \sin \te \cos \phi \nn \\
y &=& \al \sinh \varepsilon \sin \te \sin \phi \nn \\
z &=& \al \cosh \varepsilon \cos \te
\eea
It is appropriate to introduce the semi major and minor axes by the substitution,
\bea
a &=& \al \cosh \varepsilon \nn \\
b &=& \al \sinh \varepsilon
\eea
This will then reproduce Eq. \ref{elli} above. Hence we use,
\bea
x &=& b \sin \te \cos \phi \nn \\
y &=& b \sin \te \sin \phi \nn \\
z &=& a \cos \te
\eea
We also require the surface area of the ellipsoid. This can be calculated once we have the
area element $dA$ equivalent to $ r^2 d\Omega$ in spherical polar coordinates.
In the prolate spheroidal coordinate system we find,
\be
d A = h_{\te} \, h_{\phi}\, d\te\, d\phi =
( a^2 \sin^2 \te + b^2 \cos^2 \te )^{1/2} b \sin \te \, d\te \, d\phi \;\; .
\ee
where the usual $h_k$ terms are defined by the length element squared,
\be
ds^2 = h_{\varepsilon}^2 \, {d\varepsilon}^2 + h_{\te}^2 \,{d\te}^2 + h_{\phi}^2 \, {d\phi}^2 \;\; .
\ee
Now we can easily integrate the area element over all angles, $ 0 \le \te \le \pi$ and $0 \le \phi < 2 \pi$.
We will need this surface area for the moments of inertia later on.
The surface area of the ellipsoid is;
\bea
 \mbox{ Area } &=& \int_0^{2 \pi} d\phi \int_0^{\pi}
 ( a^2 \sin^2 \te + b^2 \cos^2 \te )^{1/2} b \sin \te \; d\te \nn \\
&=& 2 \pi a b \int_{-1}^{1} ( 1 - e^2 x^2 )^{1/2} dx \nn \\
&=& 4 \pi a b \int_0^e (1 - z^2 )^{1/2} dz \nn \\
&=& \frac{4 \pi a b}{e} \int_0^{\; \sin^{-1} e} \cos^2 \te  \; d\te \nn \\
\Rightarrow \mbox{ Area} &=&  2 \pi b \left( \frac{a \sin^{-1} e}{e} + b \right)
\label{area}
\eea
where in the first step we set $ x= \cos \te$, then $z = ex$, and then $z= \sin \te$.
We used the double angle formula for $\sin 2 \te = 2 \sin \te \cos \te $ and from tables \cite{tables}
we have that $ \sin^{-1} x = \cos^{-1} \sqrt{ 1 -x^2} $ so that
$ \cos ( \sin^{-1} e) = b/a$ where $e = \sqrt{ 1-b^2/a^2 }$.
At this point the derivation of the principal moments of inertia is reasonably straight forward, although
a little messy. We introduce the surface mass density $\rho = M/{\mbox{Area}}$, where the Area
is that given by Eq. \ref{area}, and define the following
principal moments;
\bea
I_1 = I_2 &=& \rho \int \int ( x^2 + z^2 ) dA \nn \\
I_3 &=& \rho \int \int ( x^2 + y^2 ) dA
\eea
where $ I_1 = I_{xx}\; $, $I_2 = I_{yy}$ and $I_3 = I_{zz}$ and
 $dA = ( a^2 \sin^2 \te + b^2 \cos^2 \te )^{1/2} b \sin \te \, d\te \,d\phi $.
To save space here we give only one derivation, the other being very similar.
\bea
I_3 &=& \rho \int \int ( x^2 + y^2 ) dA \nn \\
 &=& \rho \int_0^{2 \pi} d\phi \int_0^{\pi}  b^3 \sin^3 \te
 ( a^2 \sin^2 \te + b^2 \cos^2 \te )^{1/2} \; d\te \nn \\
 &=& 4 \pi a b^3 \rho \int_0^{\pi/2} \sin^3 \te ( 1 - e^2 \cos^2 \te )^{1/2} \; d\te \nn \\
 &=& 4 \pi a b^3 \rho \int_0^1 (1-x^2 ) ( 1 - e^2 x^2 )^{1/2} dx \nn \\
 &=& 4 \pi a b^3 \rho \int_0^e \left( 1-\frac{z^2}{e^2} \right)\left(1-z^2 \right)^{1/2}
 \frac{dz}{e} \nn \\
 &=& \frac{ 4 \pi a b^3 \rho}{e} \int_0^{ \sin^{-1} e}
 \left( \cos^2 \te - \frac{ \sin^2 2 \te }{4e^2} \right) \; d\te \nn \\
 &=& \frac{ 4 \pi a b^3 \rho}{e} \left[ \frac{1}{2} \sin^{-1}e + \frac{e}{2} \frac{b}{a}
 - \frac{1}{8 e^2} \sin^{-1} e + \frac{1}{8e} \frac{b}{a} \left( \frac{b^2}{a^2} - e^2 \right) \right]
 \label{i3}
 \eea
 After substituting for $\rho$ and some algebra we find;
 \be
I_3 = m b^2 \left[ \left( 1 - \frac{1}{4 e^2} \right) +
\frac{ \frac{b^2}{a^2} \frac{b}{2 e^2}}{ \left( \frac{a \sin^{-1} e}{e} + b \right)} \right]
\ee
It should be noted that,
\bea
\left( 1 - \frac{1}{4 e^2} \right)&=& \frac{1}{4} \left( \frac{ 3 a^2 -4b^2}{a^2 -b^2} \right) \nn \\
\rho &=& \frac{ M}{2 \pi b \left( \frac{a \sin^{-1}e}{e} + b \right) } \;\; .
\eea
As an interesting aside, one could also calculate the $I_3$ moment using rings and then
integrating from $-a \le z \le a$. This is possible because one can set $x^2 + y^2 = r^2 $ and
then from the equation of a prolate ellipsoid, Eq.(\ref{elli}), arrive at an equation for $r(z)$, $r'$
and the width of a ring $ds$ as;
\bea
r(z) &=& b \left( 1 - \frac{z^2}{a^2} \right)^{1/2} \nn \\
r'(z) = \frac{dr}{dz} &=& \frac{ bz/a^2}{ \left( 1 - \frac{z^2}{a^2} \right)^{1/2} } \nn \\
ds = \sqrt{ dr^2 + dz^2 } &=& \left[ 1 + \left( \frac{dr}{dz} \right)^2 \right]^{1/2} dz \;\; .
\eea
Therefore, with the mass of the ring as $dm_{\mbox{\small{ring}}}= \rho \, 2 \pi r \, ds $ we have;
\bea
I_3 &=& \int r^2 dm_{\mbox{\small{ring}}} \nn \\
&=& 2 \pi \rho \int r^3 ds \nn \\
&=& 4 \pi \rho \int_0^a r^3 \left[ 1 + \left( \frac{dr}{dz} \right)^2 \right]^{1/2} \, dz
\nn \\
&=& 4 \pi \rho \int_0^a b^3 \left( 1 - \frac{z^2}{a^2} \right)
\left( 1 + z^2 \frac{ (b^2 -a^2 )}{a^4} \right)^{1/2} \, dz
\eea
which after setting $z = a \cos \te$ we arrive at the third line of Eq. (\ref{i3}).\\

\noindent
Finally, we give the result for the principal axes $ I_1 = I_2 $.
\bea
I_1 &=& \rho \int \int ( x^2 + z^2 ) dA \nn \\
&=& \rho \int_0^{2 \pi} d\phi \int_0^{\pi} d\te \, ( b^2 \sin^2 \te \cos^2 \phi + a^2 \cos^2 \te )
(a^2 \sin^2 \te + b^2 \cos^2 \te )^{1/2} \; b \sin \te \nn \\
&=& 2 \pi a b \rho \int_0^{\pi/2 } \, d\te
\left[ b^2 \sin^3 \te  +
2 a^2 \sin \te \cos^2 \te \right] \left( 1 - e^2 \cos^2 \te \right)^{1/2}
\eea
After integration, substituting for $\rho$ and some algebra we get,
\be
I_1 = \frac{1}{2} m b^2 \left[ \left( 1 - \frac{1}{4e^2} + \frac{ a^2}{2 e^2 b^2} \right) +
\frac{\frac{b}{e^2}\left( \frac{b^2}{2a^2} -1 \right)}{ \left( \frac{a \sin^{-1} e}{e} +
b \right)} \right] \;\; .
\ee
These results were checked numerically on Mathematica 5.2, the numerical answers,
 for the prolate spheroid, (shell)
 using $a=0.141 $m (or $5.5$ in), $b=0.086$m ( or $3.4$ in) and $M =0.411$ Kg
were $I_1 = 0.003428 kg m^2 $ and $I_3 = 0.002138Kg m^2$.
When we use the same parameters in our exact formulae above we find exactly the same numbers, so we are
confident that the above results are correct. Mathematica did not give nicely simplified answers
so we did not attempt to use its non-numerical output.\\

\section{Appendix B}

Moments of inertia for a parabola of revolution. We also show photographs of an American football and a
Rugby ball superposed onto graph paper and curve fit using Mathematica to show how well these
respective game balls fit to a parabola of revolution and an ellipse.
It is quite clear that the American football fits the parabola of revolution
much more precisely than an ellipse. The Rugby ball is a closer fit to the ellipse shape.\\

\noindent
Consider Figure 5. of the parabola of revolution. We will show how to calculate the surface area,
and the moments of inertia along the z($e_3$) and x ($e_1$)
axes, corresponding to $I_3$ and $I_1$ respectively.

\begin{figure}
  \includegraphics[width=4.0in]{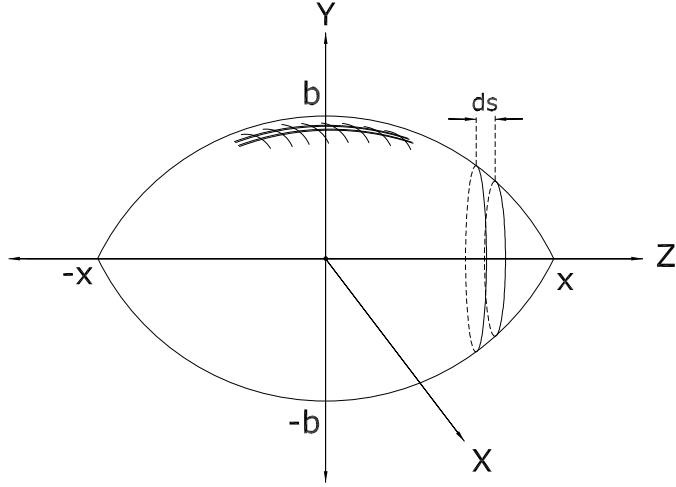}\\
  \caption{Parabola of revolution, with equation $r(z) = b - z^2 /(4a)$,
  the length of the football is $2x$ where $x=2 \sqrt{ab}$}
  \label{Figure 5.}
\end{figure}

\noindent
To calculate the moments we must first determine the surface area of the parabola of revolution.
The surface area is calculated by the simple integral $ A = \int 2 \pi r ds $ where
$ds = ( 1 + r'^2 )^{1/2} dz $ and $r' = dr/dz $. We define the semi--major axis here to be
$x = 2 \sqrt{ab}$ which will simplify the integrations considerably.
Using the parabolic equation $ r(z) = b - z^2/(4a)$
we find that the surface area of revolution is given by,

\bea
A &=& \int 2 \pi r(z) ds \nn \\
&=& 2 \pi \int_{-x}^{x} \; r [ 1 + r'^2 ]^{1/2} \; dz  \nn \\
&=& 2 \pi b^2 \int_{-x}^{x} \; \left( 1 - \frac{z^2}{x^2} \right)
\left( \frac{1}{b^2} + \frac{4 z^2}{x^4} \right)^{1/2} \; dz \\
\eea

This can easily be solved on mathematica and the result for $x = 0.141$ m and $b = 0.086$ m
is found to be
$A = 0.114938 \;\; m^2$. The calculation can be done by hand but it is very long winded and tedious.
We have not written out the full expression because it
does not lead to any great insight.\\

\noindent
The moment of inertia for the $e_3$, or long axis, is found most easily by summing over rings.
Using the area of a ring to be $ 2 \pi r ds $ and the mass of a ring
is $dm_{\mbox{\tiny ring}} = \rho 2 \pi r ds $
where $\rho = M/A$, $M = 0.411 $kg, is the total mass of the football and $A$ is the surface area
given above.

\bea
I_3 &=& \int r^2 dm \nn \\
&=& 2 \pi \rho \int r^3 ds \nn \\
&=& 2 \pi \rho  b^4 \int_{-x}^{x} \; \left( 1 - \frac{z^2}{x^2} \right)^3
\left( \frac{1}{b^2} + \frac{4 z^2}{x^4} \right)^{1/2} \; dz \\
\eea

\noindent
Making substitutions of the form $z= i x^2 \sin \theta /(2b)$ simplifies the square root term
and may allow you to solve this and the $I_1$ below by hand,
but we would not recommend it. Mathematica again comes to the rescue and we find a value of
$I_3 = 0.001982 $ kg$m^2$.\\

\noindent
There are two ways to proceed with the moment of inertia about the $e_1$ or $x$ axis. You can chop the
football into rings again and use the parallel axis theorem. Or you can directly integrate over
the surface using small elemental areas. We show the small area method below. Consider a small
area of the surface and take the mass to be $dm = \rho dA$.
Then the contribution of this small area
to the moment about $e_1$ is given by $dI_1 = \rho ( y^2 + z^2) dA$. We have taken the vertical
(or $e_2$) axis to be $y$ here. Convert to polar coordinates, using $ x = r \cos \theta$ and
$y = r \sin \theta $. Note that $x^2 + y^2 = r^2$ since there is a circular cross--section.
In the xy direction we may change to polar coordinates, $r d\theta$. In the z-direction we must use
the length $ds$ for accuracy. Therefore, an element of the surface has an
area $dA = r d\theta \; ds$ where $ds$ is defined above.

\bea
I_1 &=& 2 \rho \int_0^{2 \pi} \; d\theta \; \int_0^x \; ds \; r ( y^2 + z^2 ) \nn \\
&=& 2 \rho \int_{0}^{x}  \; \int_0^{2 \pi}  \; \left( 1 - \frac{z^2}{x^2} \right)
\left( 1 + \frac{4 b^2 z^2}{x^4} \right)^{1/2}
\left[ b^2 \left( 1 - \frac{z^2}{x^2} \right)^2 \sin^2 \theta + z^2 \right] d\theta \; dz \nn \\
&=& 2 \pi \rho \int_0^x \; \left( 1 - \frac{z^2}{x^2} \right)
\left( 1 + \frac{4 b^2 z^2}{x^4} \right)^{1/2}
\left[ b^2 \left( 1 - \frac{z^2}{x^2} \right)^2 + 2 z^2 \right] \; dz \nn \\
\eea

\noindent
For the parabola of revolution, we get $I_1 = 0.002829 $ kg $m^2$.

\begin{figure}
  \includegraphics[width=3.5in]{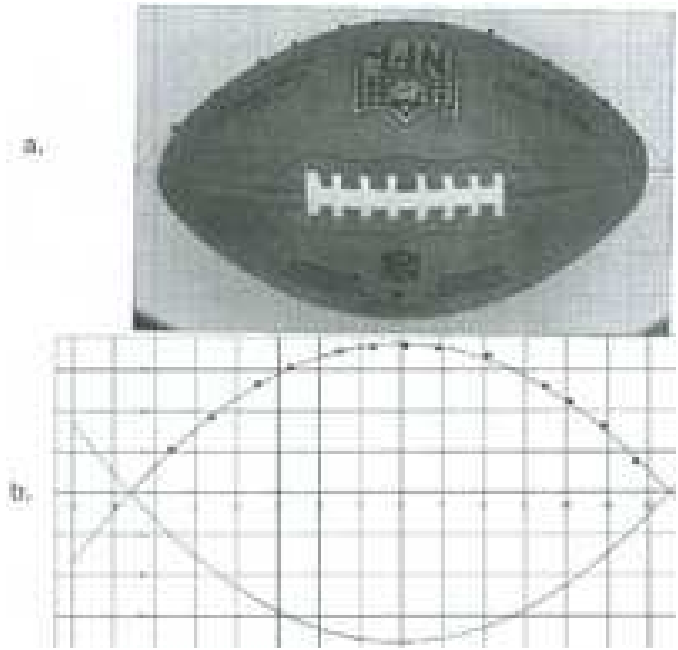}\\
  \caption{American Football Photo and curve fit. Plot(a) shows the photograph of
  the football with the outline of the curve fitting to show how well they match.
  Plot(b) shows the curve fit from Mathematica alone with the points taken from the
  original photograph of the football.}\label{Fig6}
\end{figure}

\noindent
To clarify our point, we show photographs of both an American football (pro NFL game ball) above
and a Rugby ball below. The photographs were taken with both balls on top of graph paper.
We used the outer edge of each photograph of the ball to get points to plot and curve fit with
Mathematica 5.2. The results are shown in figures 6 and 7.

\begin{figure}
  \includegraphics[width=3.5in]{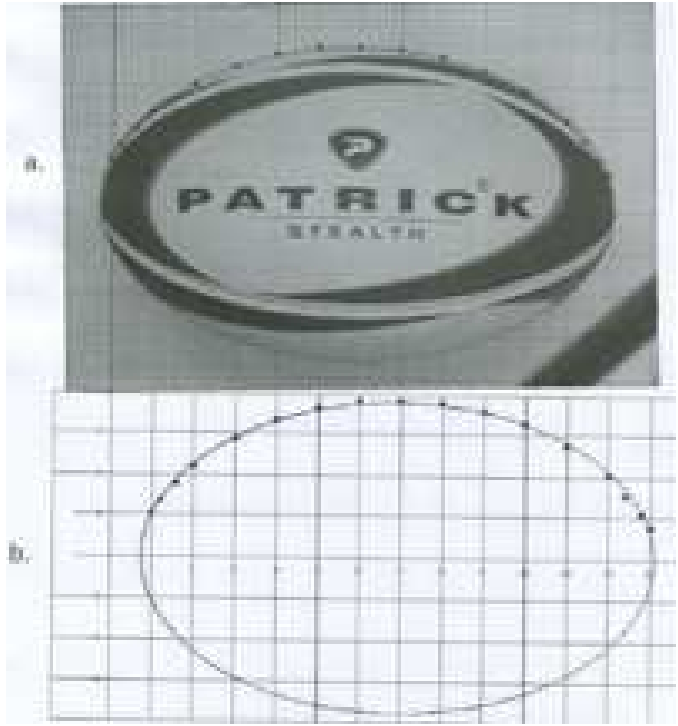}\\
  \caption{Rugby ball Photo and curve fit. Plot(a) shows the photograph of
  the rugby ball with the outline of the curve fitting to show how well they match.
  Plot(b) shows the curve fit from Mathematica alone with the points taken from the
  original photograph of the rugby ball.} \label{Fig7}
\end{figure}

\noindent
From figures 6 and 7 we see that the American football closely fits the shape of a parabola of revolution.
The rugby ball more closely fits the shape of an ellipsoid.

\newpage



\end{document}